\documentclass[12pt]{article}%
\usepackage[T2A]{fontenc}
\usepackage[cp1251]{inputenc}
\usepackage{amsmath}
\usepackage{amsfonts}
\usepackage{amssymb}
\usepackage{graphicx}
\usepackage{cite}%
\usepackage{authblk}

\begin{document}

\title{On the use of the equivalent source method for free-field calibration of an acoustic radiator in a reverberant tank}
\date{\ }

\author[1]{A.L. Virovlyansky}
\author[1,2]{M.S. Deryabin}
\affil[1]{\footnotesize Institute of Applied Physics, Russian Academy of Science, 46 Ul'yanov Street, Nizhny Novgorod, Russia, viro@appl.sci-nnov.ru}
\affil[2]{\footnotesize N.I. Lobachevsky State University of Nizhny Novgorod, 23 Prospekt Gagarina (Gagarin Avenue), Nizhny Novgorod, Russia}


\maketitle

\begin{abstract}
It is shown that the use of the equivalent source method allows one to
reconstruct the sound field excited by an acoustic radiator in free space
from the measurements of the sound filed excited by this radiator in a tank
with reflecting boundaries. The key assumption underlying the proposed
approach is that the same set of monopoles with the same strengths can
reproduce the field of the radiator both in free space and in the tank. The
monopole strengths are reconstructed from the measurements in the tank, and the
field in free space is then calculated using the known Green function. The
feasibility of the approach is demonstrated in a tank
experiment.
\end{abstract}

\newpage

\section{Introduction}

Traditional approaches to solving the problem of free-field calibration of the
acoustic radiator in a reverberant tank are based on the isolation of the
direct signals arriving at the receivers without reflection from the tank
boundaries. There are two conventional methods for eliminating the effect of
boundary echoes \cite{Bobber,Robinson1999}. One of them is to coat
the boundaries with absorbers. The second method relies on the time-gating technique for resolving the direct signals by their arrival times. However, both methods are applicable only at sufficiently high frequencies. There exist more sophisticated methods which can be used at acoustic frequencies below the limits imposed by
the echo-free time of the test tank \cite{Robinson2018,Isaev2009}.

In the present paper, we discuss an alternative approach that can be used for
a free-field calibration of an acoustic radiator in a finite-sized tank
without isolating the direct signal. To solve the problem, it is proposed to
apply the well-known equivalent source method \cite{Koopmann1989,
Bobrovnitskii95,Johnson1998,Zhang2009,Gounot2011,Lee2017}. The sound field excited by the radiator is represented as a weighted superposition of fields
excited by a set of acoustic monopoles called the equivalent sources. Our main
assumption is that, under certain conditions set forth below, the radiator
field in free space and in the tank can be reproduced by the same set of
monopoles with the same strengths (amplitudes). The equivalent source
strengths are reconstructed from the tank measurements. To solve this inverse
problem, one needs to know the Green function of the Helmholtz equation in the
tank. The required values of this function are obtained using the procedure
that we call the tank calibration. It consists in placing the etalon monopole
in the equivalent source positions and recording the emitted signals by all
the receivers. The desired radiator field in free space is then readily evaluated using the
reconstructed source strengths and the known Green function.

The paper is organized as follows. Section \ref{sec:esm} outlines the main
idea of our approach. The conditions under which it can be used are formulated
and the procedure for solving the inverse problem is proposed. The tank
experiment conducted to demonstrate the feasibility of this approach is
described in Sec. \ref{sec:experiment}. In section  \ref{sec:6and8}, it is
shown, using the experimental data, that the same set of equivalent sources
can reproduce the field of a single monopole both in free space and in a tank.
Sec. \ref{sec:dipole} describes the reconstruction of the fields of acoustic
dipole and quadrupole in free space from the measurement of their fields in
the tank. The results of the paper are summarized in Sec.
\ref{sec:conclusion}.

\section{Equivalent source method \label{sec:esm}}

\subsection{Free space \label{sub:free}}

Consider a radiator exciting a monochromatic wave field at the carrier
frequency $f$ in free space with the sound speed $c$. In the framework of the
equivalent source method, the complex field amplitude $u$ at an arbitrary
observation point $\mathbf{r}$ is represented as a superposition of fields of
acoustic monopoles located at $N$ points $\mathbf{r}_{1},\ldots,\mathbf{r}%
_{N}$:%
\begin{equation}
u\left(  \mathbf{r}\right)  =\sum_{n=1}^{N}G\left(  \mathbf{r},\mathbf{r}%
_{n}\right)  A_{n}, \label{uGA}%
\end{equation}
where
\begin{equation}
G\left(  \mathbf{r},\mathbf{r}_{n}\right)  =\frac{1}{\left\vert \mathbf{r}%
-\mathbf{r}_{n}\right\vert }e^{ik\left\vert \mathbf{r}-\mathbf{r}%
_{n}\right\vert } \label{G-free}%
\end{equation}
is the Green function in free space, $k=2\pi f/c$ is the wave number, and
$A_{n}$ is the $n$-th source strength. Here, and in what follows, the time
factor $e^{-2\pi ift}$ is omitted.

To find the source strengths one should know the values of the field amplitude
$u$ at the collocation points $\pmb{\rho}_{1},\ldots,\pmb{\rho}_{M}$
covering the measurement surface $S_{M}$ around the radiator (Fig. 1). The
values of $A_{n}$ should be found from the system of linear equations, in
matix form,
\begin{equation}
\mathbf{u=GA}, \label{uGA-matr}%
\end{equation}
where $\mathbf{u}=\left[  u\left(\pmb{\rho}_{1}\right)  ,\ldots,u\left(\pmb{\rho}
_{M}\right)  \right]  ^{T}$ is the vector of field values measured at the
collocation points, symbol T means transpose operation, $\mathbf{G}$
is the matrix of size $M\times N$ with elements $G\left(  \pmb{\rho}_{m}%
,\pmb{r}_{n}\right)  $, $\mathbf{A}=[A_{1},\ldots,A_{N}]^{T}$ is the vector
of unknown source strengths. In practice, the system of equations
(\ref{uGA-matr}) is typically underdetermined "since there are often more
waves in the model than measurement points" \cite{Gerstoft2017}.

\begin{figure*}
\begin{center}
\includegraphics[width=0.9\linewidth]{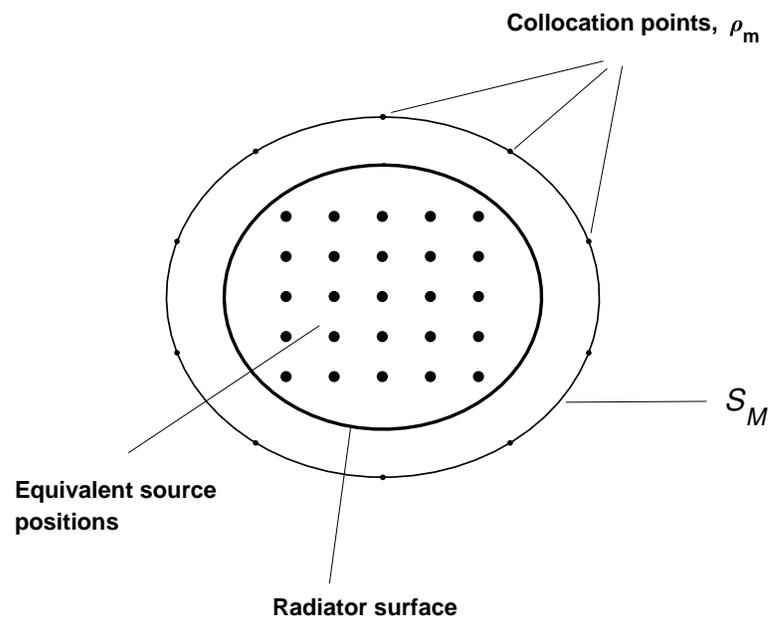}
\caption{\label{fig1}{Sketch illustrating the equivalent source method. Equivalent source positions are located inside the area occupied by the radiator.}}
\end{center}
\end{figure*}

There are well-known methods for solving such systems of equations
\cite{Semenova2005,Gerstoft2017}. One of them will be applied in this paper.
It is based on the use of the singular value decomposition of matrix
$\mathbf{G}$ \cite{Golub},%
\[
\mathbf{G}=\sum_{l=1}^{L}\gamma_{l}\pmb{\xi}_{l}\pmb{\eta}_{l}^{H},
\]
where $\gamma_{l}$ are singular values numbered in descending order,
$\pmb{\xi}_{l}$ and $\pmb{\eta}_{l}$ are singular vectors, and the
symbol $H$ denotes Hermitian conjugation. Neglecting the terms corresponding
to small singular values, we obtain an estimate of the vector $\mathbf{A}$
\begin{equation}
\mathbf{\hat{A}=}\sum_{l=1}^{L_{1}}\frac{1}{\gamma_{l}}\left(  \pmb{\xi
}_{l}^{H}\mathbf{u}\right)  \pmb{\eta}_{l}, \label{At}%
\end{equation}
where $L_{1}\leq L$ is the number of singular values taken into account.
Substitution of this estimate into the right-hand side of Eq. (\ref{uGA-matr})
yields%
\begin{equation}
\mathbf{\hat{u}=G\hat{A}=}\sum_{l=1}^{L_{1}}\left(  \pmb{\xi}_{l}%
^{H}\mathbf{u}\right)  \pmb{\xi}_{l}\mathbf{.} \label{ut}%
\end{equation}
Parameter%
\begin{equation}
\mu=\frac{\left\vert \left\vert \mathbf{u}-\mathbf{\hat{u}}\right\vert
\right\vert }{\left\vert \left\vert \mathbf{u}\right\vert \right\vert },
\label{mu}%
\end{equation}
where the symbol $\left\vert \left\vert \ldots\right\vert \right\vert $
denotes the Euclidean norm of the vector, quantitatively characterizes the
accuracy with which the total field of equivalent sources approximates the
radiated field at the collocation points. If the collocation points densely
cover the measurement surface $S_{M}$ and $\mu\ll1$, then outside the area
bounded by this surface, the equivalent sources well reproduce the field of
the radiator.

A weak point of the equivalent source method is the lack of general rules for
choosing the necessary number of equivalent sources and their positions
\cite{Bobrovnitskii95,Lee2017}. In practice, the problem of selecting the
points $\mathbf{r}_{n}$ and $\pmb{\rho}_{m}$ is usually solved empirically,
that is, through trial-and-error.

It should also be noted that due to the underdeterminedness of the system
(\ref{uGA-matr}), its solution is ambiguous. Generally, there are many
estimates $\mathbf{\hat{A}}$ (obtained using the relation (\ref{At}) or found
by other methods) that the vector $\mathbf{\hat{u}=G\hat{A}}$ approximates
$\mathbf{u}$ with high accuracy. It turns out that the 
estimates $\mathbf{\hat{A}}$, giving close $\mathbf{\hat{u}}$, may be quite
different \cite{Gerstoft2017}.

\subsection{Tank \label{sub:esm-tank}}

Consider a situation when the problem of approximating the radiator field in
free space by a superposition of fields of acoustic monopoles is solved. It
means that the positions $\mathbf{r}_{n}$ of the equivalent sources and the
source strengths $A_{n}$ are chosen in such a way that the sum on the
right-hand side of Eq. (\ref{uGA}) with high accuracy approximates the
radiator field outside some area $\sigma_{M}$. The latter can be a volume
bounded by the measurement surface $S_{M}$, or a part of this volume.

Let us imagine that a reflecting boundary is placed in free space outside the
area $\sigma_{M}$. The radiator and the equivalent sources excite practically
equal waves incident on this boundary. It is clear that after reflection from
the same boundary, the waves will remain equal. Moreover, if the radiator is
sufficiently small, its field and the total field of equivalent sources will
remain approximately the same, even in the presence of several boundaries,
when the waves experience multiple reflections. The smallness of the radiator
is required because the waves reflected from the boundaries 'do not notice'
the monopole equivalent sources, but they are scattered on the radiator of
finite size. It is natural to expect that with a sufficiently small radiator
size, the effect of this scattering will be insignificant.

Based on the above elementary reasoning, we will assume that a set of
equivalent sources simulating the fields of a radiator in free space can be
used to represent the field of this radiator in a tank with reflecting
boundaries. In both cases, the sources are located within the same area
$\sigma_{M}$. In the tank, Eq. (\ref{uGA}) translates to%
\begin{equation}
\tilde{u}\left(  \mathbf{r}\right)  =\sum_{n=1}^{N}\tilde{G}\left(
\mathbf{r},\mathbf{r}_{n}\right)  A_{n}, \label{uGAt}%
\end{equation}
where $\tilde{G}\left(  \mathbf{r},\mathbf{r}_{n}\right)  $ is the field of
the acoustic monopole (Green function) in the tank, $\mathbf{r}_{n}$ and
$A_{n}$ are, respectively, \textbf{the same} positions and strengths of the equivalent
sources as in Eq. (\ref{uGA}). We assume that this representation of the
radiator field in the tank is valid under the following conditions.

(i) The sound speed in the tank is constant and equal to its value in free space.

(ii) The area $\sigma_{M}$ can fit entirely in the tank. Equation (\ref{uGAt})
describes the sound field only at points $\mathbf{r}$ located outside
$\sigma_{M}$.

(iii) The radiator in the tank works in the same way as in free space: each
point of its surface in both cases oscillates according to the same law.

(iv) The radiator sizes are small compared to the tank sizes.

Our idea of restoring the radiator field in free space from the measurements in a
tank is based on the conjecture that the opposite is also true. We suppose
that if a set of equivalent sources reproduces the radiator field in the tank,
then the field of the same radiator in free space is well approximated by the
same set of equivalent sources, that is, by the monopoles with the same
$\mathbf{r}_{n}$ and $A_{n}$. As in free space, the field in the tank should
be measured at the collocation points $\pmb{\rho}_{1},\ldots,\pmb{\rho
}_{M}$. Field amplitudes at these points form the vector $\mathbf{\tilde{u}%
}=\left[  \tilde{u}\left(  \pmb{\rho}_{1}\right)  ,\ldots,\tilde{u}\left(
\pmb{\rho}_{M}\right)  \right]  ^{T}$. From Eq. (\ref{uGAt}) we find a
system of linear equation similar to (\ref{uGA-matr}):%
\begin{equation}
\mathbf{\tilde{u}=\tilde{G}A}, \label{uGAt-matr}%
\end{equation}
where $\mathbf{\tilde{G}}$ is a matrix of size $M\times N$ with the elements
$\tilde{G}\left(  \pmb{\rho}_{m},\mathbf{r}_{n}\right)  $, $\mathbf{A}$ is
a vector of the source strengths. In contrast to the free space, the Green
function $\tilde{G}\left(  \pmb{\rho},\mathbf{r}\right)  $ in the tank is
unknown. Therefore, the quantities of $\tilde{G}\left(  \pmb{\rho}%
_{m},\mathbf{r}_{n}\right)  $ for each pair $\left(  \pmb{\rho}%
_{m},\mathbf{r}_{n}\right)  $ must be measured. To this end, one should use a
reference acoustic monopole, alternately placed at each of the points
$\mathbf{r}_{n}$. The complex amplitude of the tone emitted from the point
$\mathbf{r}_{n}$ and received at the point $\pmb{\rho}_{m}$ is proportional
to $\tilde{G}\left(  \pmb{\rho}_{m},\mathbf{r}_{n}\right)  $. The procedure
of measuring the coefficients $\tilde{G}\left(  \pmb{\rho}_{m}%
,\mathbf{r}_{n}\right)  $ we call the tank calibration.

The desired radiator field in free space is calculated by the formula
(\ref{uGA}) after substituting into it the source strengths $A_{n}$, found
from the solution of system (\ref{uGAt-matr}).

\section{Tank experiment \label{sec:experiment}}

To test the feasibility of the proposed approach, as well as the validity of the
assumptions underlying it, a tank experiment was conducted. For its
implementation was used the experimental setup developed on the basis of a
measurement system (Ultrasound Measurement System Control Centre produced by
Precision Acoustics company) which includes an organic-glass container of
dimensions 1x1x1 m (Fig. 2) \cite{Gurbatov2019}. This container played the role of our tank. It
was filled with clean degassed and deionized water with a specific resistance
of at least 18 MOhm*cm, which was produced by a DM-4B membrane-type distiller
unit. The water temperature was monitored with a thermometer and was measured
at $24\pm0.1^{\circ}$C in the experiment.

\begin{figure*}
\begin{center}
\includegraphics[
width=0.7\linewidth]{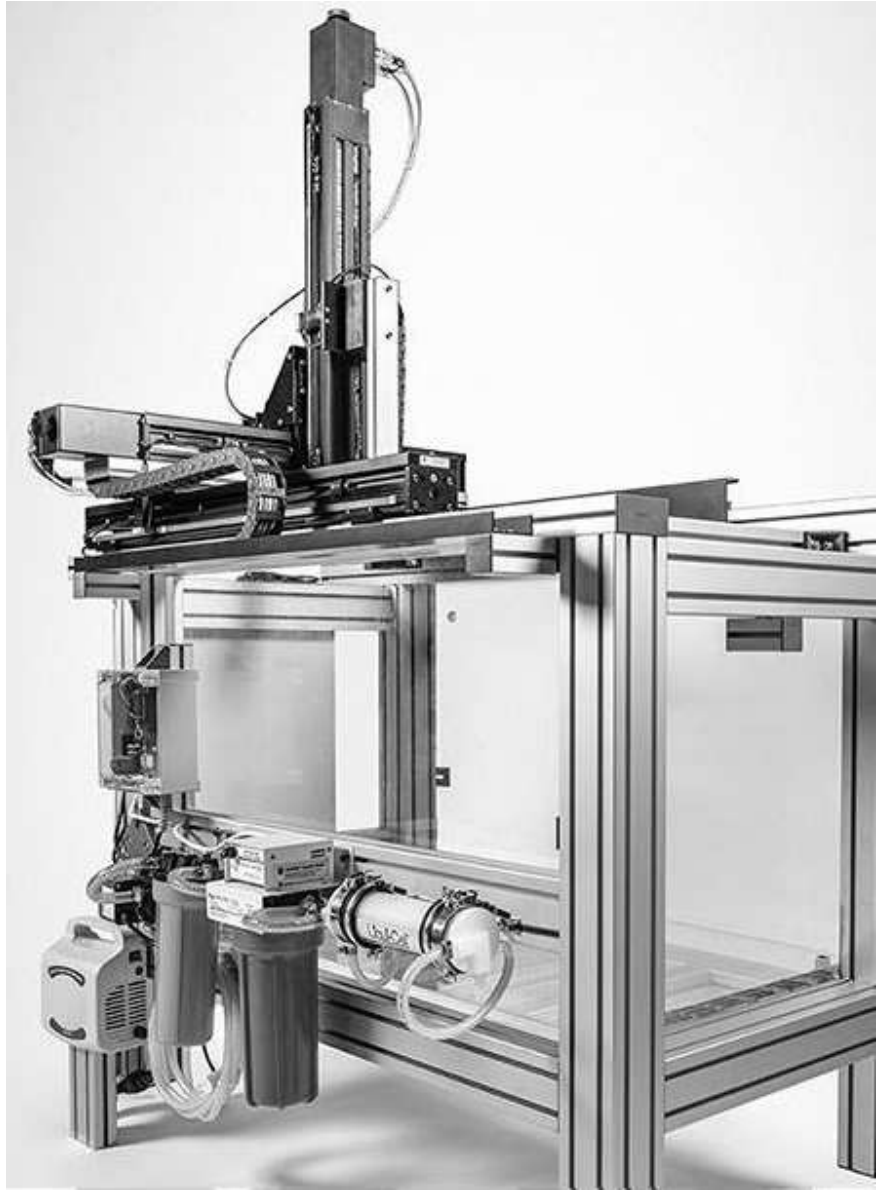}
\end{center}
\caption{Experimental setup.}%
\label{fig2}%
\end{figure*}

Two reversible hydrophones B \& K8103 were used to tank calibration. Each of
them has an operating band of up to 180 kHz, with dimensions of $10\times16$
mm. Both hydrophones were attached to the manipulators through metal tubes 10
mm in diameter with the length of the submerged part up to 0.5 m. The absolute
accuracy of the manipulator-assisted motion amounts to 6 $\mu$m.

The positions of the hydrophones in the tank will be described using the
Cartesian coordinate system $(x,y,z)$ with the $y$ axis directed vertically upwards.

One of the hydrophones was used as a sound source. The radiation was carried
out at the carrier frequency $f$ = 13.7 kHz, which corresponds to the
wavelength $\lambda$ = 10.8 cm. Since the hydrophone dimensions indicated 
above are small
compared to $\lambda$, this sound source was an acoustic monopole. It was alternately placed into
points $\mathbf{r}_{n}$ forming a $7\times7\times7$ cube, whose center
coincided with the origin. Figure 3 shows the cross section of this cube in
the $x=0$ plane. The distance between the nearest points of the cube is 1 cm,
that is, about 0.1 $\lambda$.

The second hydrophone was used as a sound receiver. It was located in five
positions $\pmb{\rho}_{1}$, \ldots,$\pmb{\rho}_{5}$, shown in Fig. 3 by
bold dots. The step size between the neighboring receiver positions is 5 cm.
They are placed along a straight line parallel to the y axis. The line is
shifted at the distance $h$ = 38.5 cm from the origin.

\begin{figure*}
\begin{center}
\includegraphics[
width=0.9\linewidth ]{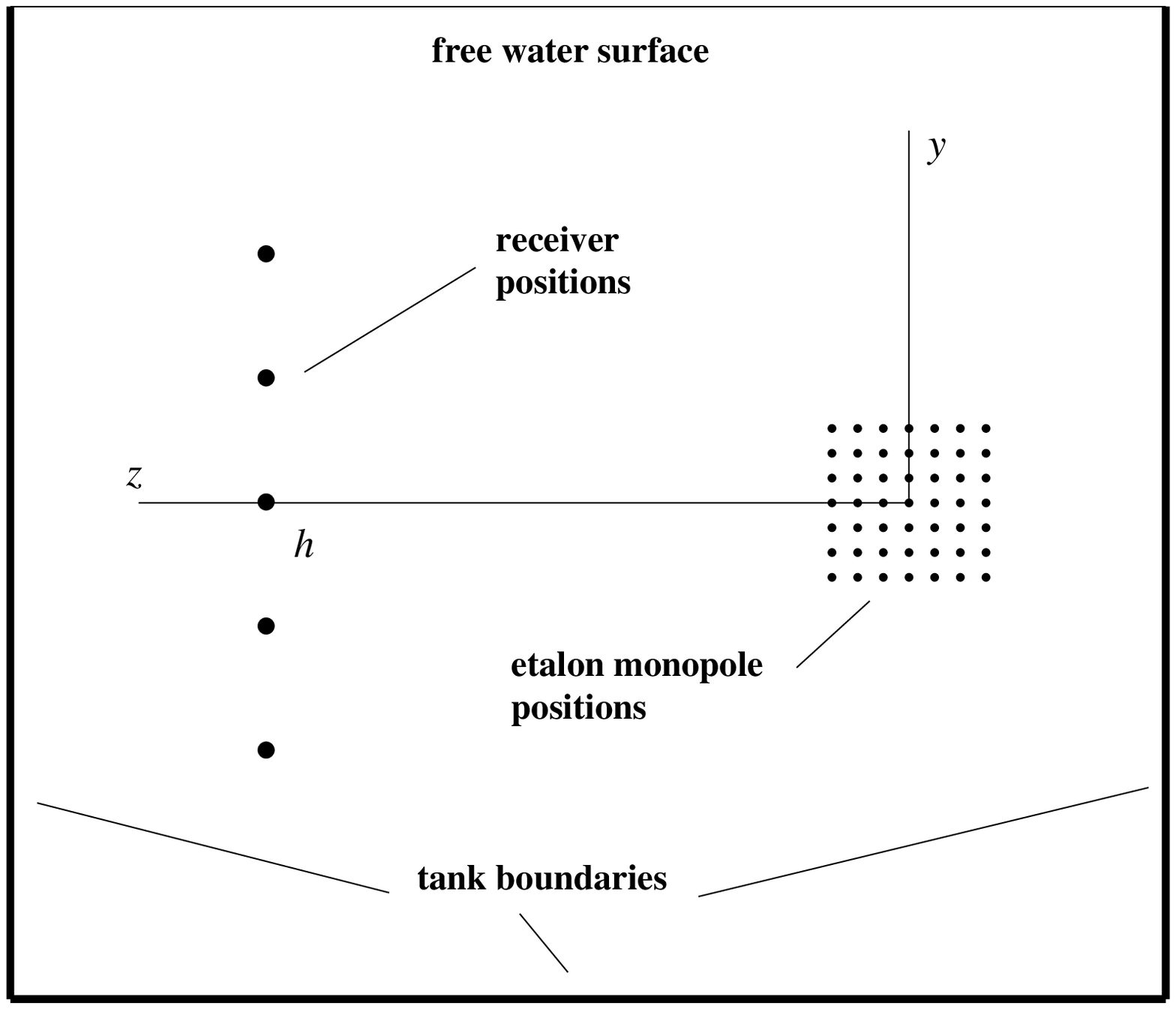}
\end{center}
\caption{Section of the tank in the plane x = 0. Location of the receiver positions and points at which the etalon monopole was placed during the tank calibration. }%
\label{fig3}%
\end{figure*}

The measurements were performed as follows. While the receiving hydrophone was
located at one the points $\pmb{\rho}_{1}$, \ldots,$\pmb{\rho}_{5}$, the
hydrophone playing the role of the source was placed in turn at all
$7^{3}=343$ points $\mathbf{r}_{n}$ forming the cube. From each point it
emitted the same tonal signal at frequency $f$, which was recorded by the
receiver. Registration began with a delay of $\tau$ = 2000 ms after the start of radiation. During the time $\tau$, all the transients associated with the reflections 
from the boundaries and
resonant behavior of the source ended and the received signal was an almost
perfect sinusoid. The signal emitted from the point $\mathbf{r}_{n}$ and
recorded at the point $\pmb{\rho}_{m}$ was a piece of sinusoid $g_{m,n}%
\sin\left(  2\pi ft-\phi_{m,n}\right)  $ including of about 27 periods of the
carrier frequency. The time $t$ for each signal was reckoned from the start of
the radiation $t$ = 0. The complex demodulate of this signal $g_{m,n}%
\exp\left(  i\phi_{\mu,n}\right)  $ is equal to $C\tilde{G}\left(
\pmb{\rho}_{m},\mathbf{r}_{n}\right)  $, where $C$ is some constant complex
factor that is the same for all pairs $(m,n)$. In what follows, we will
compare only linear combinations of $\tilde{G}\left(  \pmb{\rho}%
_{m},\mathbf{r}_{n}\right)  $ among themselves and, therefore, the factor $C$
will be omitted.

In the course of the experiment, $5\times343=\allowbreak1715$ complex values
of the Green's function in the tank, $\tilde{G}\left(  \pmb{\rho}%
_{m},\mathbf{r}_{n}\right)  $, were measured. These data were used to test the
approach under consideration. The results are presented in the next sections.

\section{Approximation of the acoustic monopole field using 6 and 8 equivalent
sources \label{sec:6and8}}

We begin by checking the assumption that under conditions (i) - (iv) listed in
Sec. \ref{sub:esm-tank} the same set of equivalent sources (acoustic
monopoles) reproducing the field of a given radiator in free space reproduces
the field of this radiator in the tank. To this end, consider two simple
examples in which the source strengths $A_{n}$ in Eq. (\ref{uGA}) are found analytically.

In both examples, the role of the radiator is played by an acoustic monopole
located at the point $\mathbf{r}=0$ and exciting the field $G\left(
\mathbf{r},0\right)  =\exp\left(  ikr\right)  /r$, where $r=\left\vert
\mathbf{r}\right\vert $. Further, this monopole will be called central. The
function $G\left(  \mathbf{r},0\right)  $ will be approximated by a
superposition of fields of equivalent sources representing exactly the same monopoles.

\begin{figure*}
\begin{center}
\includegraphics[
width=0.7\linewidth]{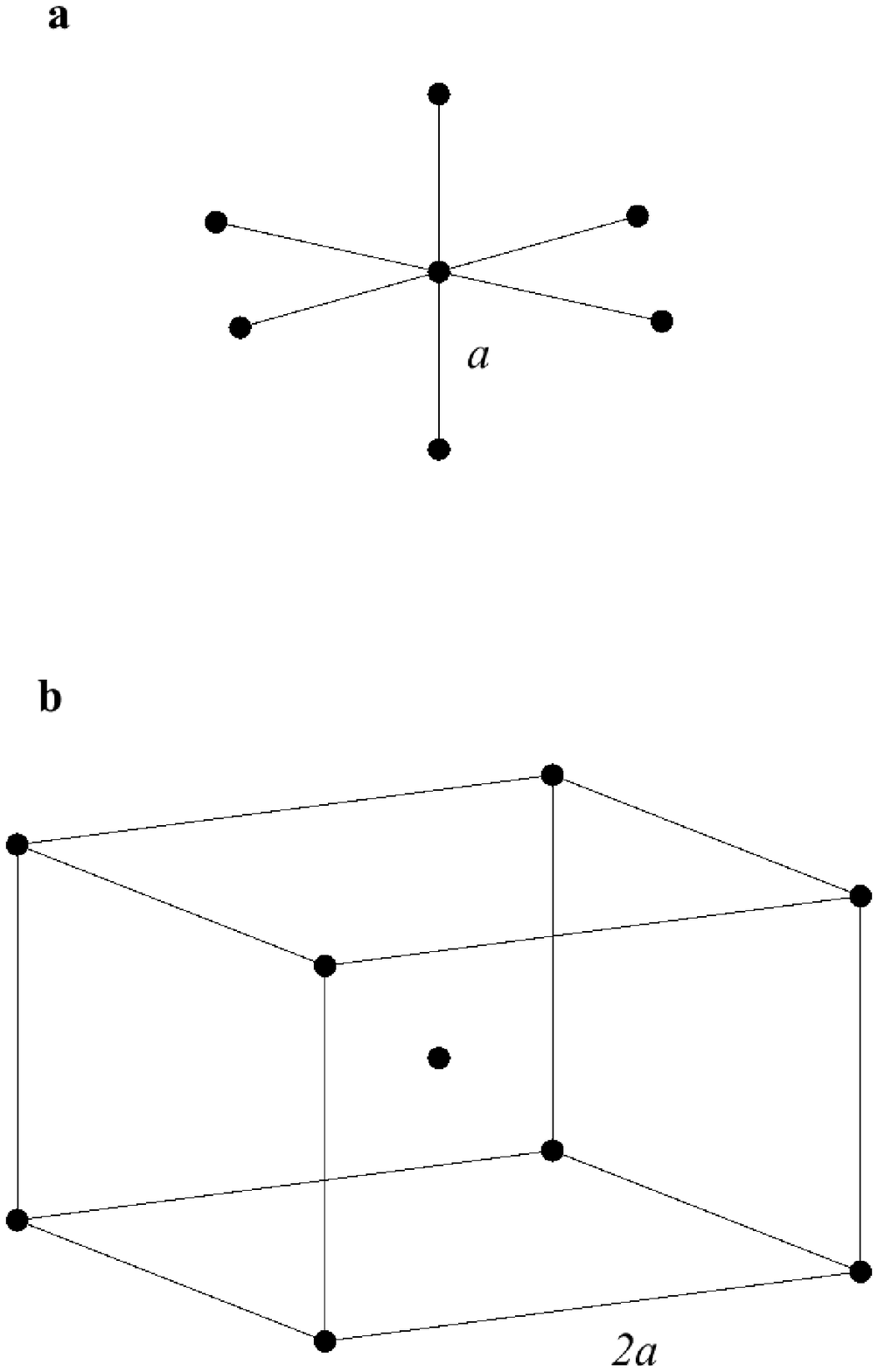}
\end{center}
\caption{The central monopole surrounded by 6 (a) and 8 (b) monopoles, considered as equivalent sources. }%
\label{fig4}%
\end{figure*}

In the first example, we consider 6 equivalent sources located at the points
$\mathbf{q}_{1,2}=\left(  \pm a,0,0\right)$, $\mathbf{q}_{3,4}=\left(
0,\pm a,0\right)$, $\mathbf{q}_{5,6}=\left(  0,0,\pm a\right)$,
that is, shifted relative to the central monopole at distances of $\pm a$
along each of the coordinate axes (Fig. 4a). Let us show that under the
condition
\begin{equation}
ka < 1 \label{ka}%
\end{equation}
the equivalent source strengths can be chosen in such a way that in the far
field, that is, at $r\gg a^{2}/\lambda$, they approximate the central monopole
field $G\left(  \mathbf{r},\mathbf{0}\right)  $ with good accuracy. From
symmetry considerations, it is clear that all the source strengths should be
equal to the same value, which we denote by $B_{6}$. It should be found from
the condition that in the far zone the equality%
\begin{equation}
G\left(  \mathbf{r},0\right)  =B_{6}\sum_{n=1}^{6}G\left(  \mathbf{r}%
,\mathbf{q}_{n}\right)  \label{G-6}%
\end{equation}
holds. In the far zone%
\[
\sum_{n=1}^{6}G\left(  \mathbf{r},\mathbf{q}_{n}\right)  =\frac{e^{ikr}}%
{r}\sum_{n=1}^{6}e^{-ik\pmb{\nu}\mathbf{q}_{n}~}%
\]%
\begin{equation}
=2\frac{e^{ikr}}{r}\left[  \cos\left(  ka\nu_{x}\right)  +\cos\left(
ka\nu_{y}\right)  +\cos\left(  ka\nu_{y}\right)  \right]  , \label{u-6}%
\end{equation}
where $\pmb{\nu}=\mathbf{r/}r$ is the unit vector whose projections onto
the axes of the Cartesian coordinate system $\left(  x,y,z\right)  $ we denote
$\nu_{x}$, $\nu_{y}$, and $\nu_{z}$, respectively. When the condition
(\ref{ka}) is met, the cosine arguments in the last expression are relatively small and
we can use the approximation
\begin{equation}
\cos\alpha\simeq1-\alpha^{2}/2. \label{cos}%
\end{equation}
Since $\nu_{x}^{2}+\nu_{y}^{2}+\nu_{z}^{2}=1$, we find%
\[
\sum_{n=1}^{6}G\left(  \mathbf{r},\mathbf{q}_{n}\right)  =\frac{e^{ikr}}%
{r}\left[  6-\left(  ka\right)  ^{2}\right]  .
\]
It follows that the equality (\ref{G-6}) holds when%
\begin{equation}
B_{6}=\frac{1}{6}+\frac{\left(  ka\right)  ^{2}}{36}. \label{A6}%
\end{equation}

Our assumption is that Eq. (\ref{G-6}) in the tank translates to%
\begin{equation}
\tilde{G}\left(  \mathbf{r},0\right)  =B_{6}\sum_{n=1}^{6}\tilde{G}\left(
\mathbf{r},\mathbf{q}_{n}\right)  , \label{Gt-6}%
\end{equation}
where $B_{6}$ is the same coefficient defined by Eq. (\ref{A6}) as in free
space, but the Green function $\tilde{G}\left(  \mathbf{r},\mathbf{q}%
_{n}\right)  $, which accounts for multiple reflections from the boundaries,
is quite different from $G\left(  \mathbf{r},\mathbf{q}_{n}\right)  $.

To verify this statement, we use the quantities $\mathbf{\tilde{G}}\left(
\pmb{\rho}_{m},\mathbf{r}_{n}\right)  $ measured in the tank experiment
described in the previous section. As the position of the central monopole
$\mathbf{r}_{\ast}$, take one of the points $r_{n}$ that does not lie on the
boundary of the cube. In this case, the roles of $\mathbf{q}_{1}%
,\ldots,\mathbf{q}_{6}$ will play cube points shifted relative to
$\mathbf{r}_{\ast}$ along each of the coordinate axes by $\pm a$, where $a$ =
1 cm. So we have $ka=0.58$ which satisfies the condition (\ref{ka}). Substituting this value into Eq. (\ref{A6})
yields%
\begin{equation}
B_{6}=0.176. \label{A-6}%
\end{equation}
Although our value of $ka$
is not very small, the difference between the left and
right sides of Eq. (\ref{G-6}) with this value of $B_{6}$ at distances from the central monopole
exceeding 5 cm is less than 0.5 \%. Therefore, a ball with a radius of 5 cm,
centered at $\mathbf{r}_{\ast}$, can be considered as the area $\sigma_{M}$
introduced in Sec. \ref{sub:esm-tank}. The points $\pmb{\rho}_{1}%
,\ldots,\pmb{\rho}_{5}$ lie outside this ball.

In Fig. 5, circles show the real (top) and imaginary (bottom) parts of the
complex pressure amplitudes $p$ (in arbitrary units) recorded at all five
receiver positions after the etalon monopole has been placed at $\mathbf{r}%
_{\ast}$. Asterisks show the real and imaginary parts of the pressures
obtained by summing the fields of the etalon monopole emitted from the six
points closest to $\mathbf{r}_{\ast}$ with the weighting factor (\ref{A-6}).
In other words, the asterisks present the field amplitude values calculated by
the formula (\ref{Gt-6}). Similar results are obtained with other choices of
the central monopole position $\mathbf{r}_{\ast}$ (not shown). Thus,
experimental data confirm the applicability of Eq. (\ref{Gt-6}).

\begin{figure*}
\begin{center}
\includegraphics[
width=0.7\linewidth]{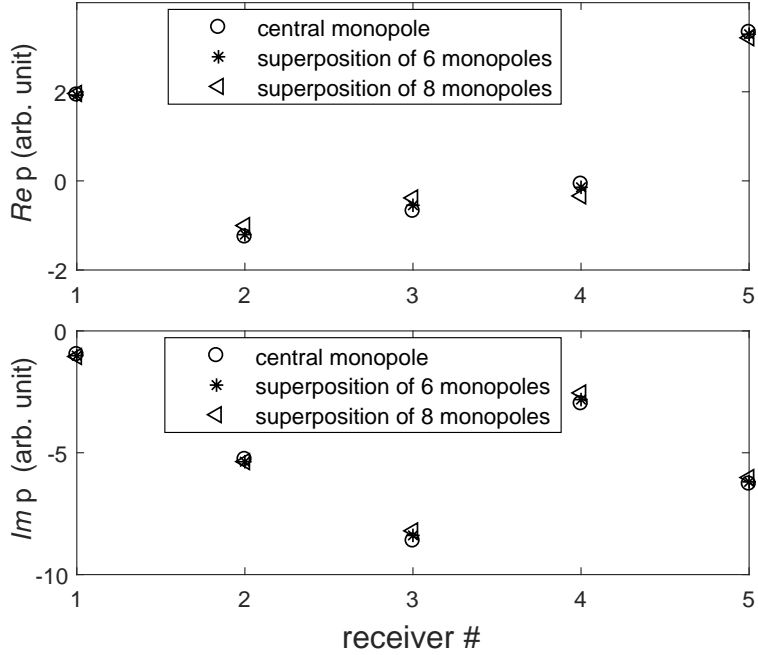}
\end{center}
\caption{Real (top) and imaginary (bottom) parts of pressure amplitudes at receiving points. Circles: signal of the central monopole. Asterisks: the sum of signals from 6 monopoles with weight $B_6$. Triangles: the sum of signals from 8 monopoles with weight $B_8$. }%
\label{fig5}%
\end{figure*}

As a second example, consider the approximation of the
acoustic monopole field by the fields of eight equivalent sources located
at the points $\mathbf{p}_{1,\ldots,8}=\left(  \pm a,\pm a,\pm a\right)%
$, that is, at the vertices of a cube with an edge of length $2a$ (Fig. 4b).
The monopole playing the role of a radiator is located in the center of this
cube. As in the previous example, the strengths of all the equivalent sources
should be the same. In this example, Eq. (\ref{G-6}) turns to%
\begin{equation}
G\left(  \mathbf{r},0\right)  =B_{8}\sum_{m=1}^{8}G\left(  \mathbf{r}%
,\mathbf{p}_{m}\right)  . \label{G-8}%
\end{equation}
In the far zone
\[
\sum_{m=1}^{8}G\left(  \mathbf{r},\mathbf{p}_{m}\right)  =\frac{e^{ikr}}%
{r}\sum_{m=1}^{8}e^{-ik\mathbf{np}_{m}}%
\]%
\[
=2\frac{e^{ikr}}{r}\left[  \cos\left(  ka\left(  n_{x}+n_{y}+n_{z}\right)
\right)  +\cos\left(  ka\left(  -n_{x}+n_{y}+n_{z}\right)  \right)  \right.
\]%
\[
\left.  +\cos\left(  ka\left(  n_{x}-n_{y}+n_{z}\right)  \right)  +\cos\left(
ka\left(  -n_{x}-n_{y}+n_{z}\right)  \right)  \right]  .
\]
Assuming that the condition (\ref{ka}) is fulfilled and again using the
approximation (\ref{cos}), we find%

\begin{equation}
\sum_{m=1}^{8}G\left(  \mathbf{r},\mathbf{p}_{m}\right)  =\frac{e^{ikr}}%
{r}\left[  8-4\left(  ka\right)  ^{2}\right]  . \label{u8}%
\end{equation}
The equality (\ref{G-8}) holds when%
\begin{equation}
B_{8}=\frac{1}{8}+\frac{\left(  ka\right)  ^{2}}{16}. \label{A8}%
\end{equation}
In the tank, the analogue of Eq. (\ref{G-8}) has the form (cf. Eq. (\ref{Gt-6}))%

\begin{equation}
\tilde{G}\left(  \mathbf{r},0\right)  =B_{8}\sum_{n=1}^{8}\tilde{G}\left(
\mathbf{r},\mathbf{p}_{n}\right)  . \label{Gt-8}%
\end{equation}

Substituting $ka=0.58$ in Eq. (\ref{A8}) yields%
\begin{equation}
B_{8}=0.149. \label{A-8}%
\end{equation}
As $\sigma_{M}$ we can take the same ball as in the previous example. Outside
this ball in free space, the difference between the left and right sides of
Eq. (\ref{G-8}) does not exceed 3\%.

To test Eq. (\ref{Gt-8}), we again use our experimental data. The triangles in
Fig. 5 present the result of summing the fields emitted from 8 points located
at the vertices of the cube shown in Fig. 4b. Fields are summed with the
weight factor (\ref{A-8}). The cube center is the same point $\mathbf{r}_{\ast
}$, which was considered in the previous example. Similar results were
obtained for other points of the cube selected as $\mathbf{r}_{\ast}$ (not shown).

\section{Reconstruction of sound fields excited by a dipole and quadrupole
\label{sec:dipole}}

Let us use the data of our measurements to verify the method of restoring
the radiator field in free space, formulated in Sec. \ref{sub:esm-tank}. In
the central part of the cube of size $7\times7\times7$ formed by the points in
which the etalon monopole was placed, we select a small cube of size
$3\times3\times3$ so that both cubes are centered at point $\mathbf{r}=0$. The
points of the big cube that are not included in the small one will be considered
as positions of the equivalent sources and denoted by the symbols
$\mathbf{\tilde{r}}_{n}$, $n=1,\ldots,316$.

We will use the superpositions of signals emitted from the points of a small
cube to simulate the fields of the dipole and quadrupole sound sources. Take
two points of a small cube, which lie on the $y$-axis and in Fig. 6 are
marked with $+$ and $-$. Denote their positions as $\mathbf{r}_{+}$ and
$\mathbf{r}_{-}$. The difference of the fields emitted by the etalon monopole
from these points, that is, $\tilde
{u}\left(  \mathbf{r}\right)  =\tilde{G}\left(  \mathbf{r},\mathbf{r}%
_{+}\right)  -\tilde{G}\left(  \mathbf{r},\mathbf{r}_{-}\right)  $, represents
the field of the acoustic dipole in the tank. In the process of tank
calibration, the values of $\tilde{G}\left(  \pmb{\rho}_{m},\mathbf{r}%
_{\pm}\right)  $, $m=1,\ldots,5$, were measured. This allows us to find the
complex amplitudes of the dipole field at the reception points $\mathbf{\tilde
{u}}=\left[  \tilde{u}\left(  \pmb{\rho}_{1}\right)  ,\ldots,\tilde
{u}\left(  \pmb{\rho}_{5}\right)  \right]  ^{T}$. This vector represents
the right-hand side of the system (\ref{uGAt-matr}), which is to be solved for
unknown source strengths $A_{n}$. The elements of matrix $\mathbf{\tilde{G}}$
were actually measured during the tank calibration.

\begin{figure*}
\begin{center}
\includegraphics[
width=0.7\linewidth]{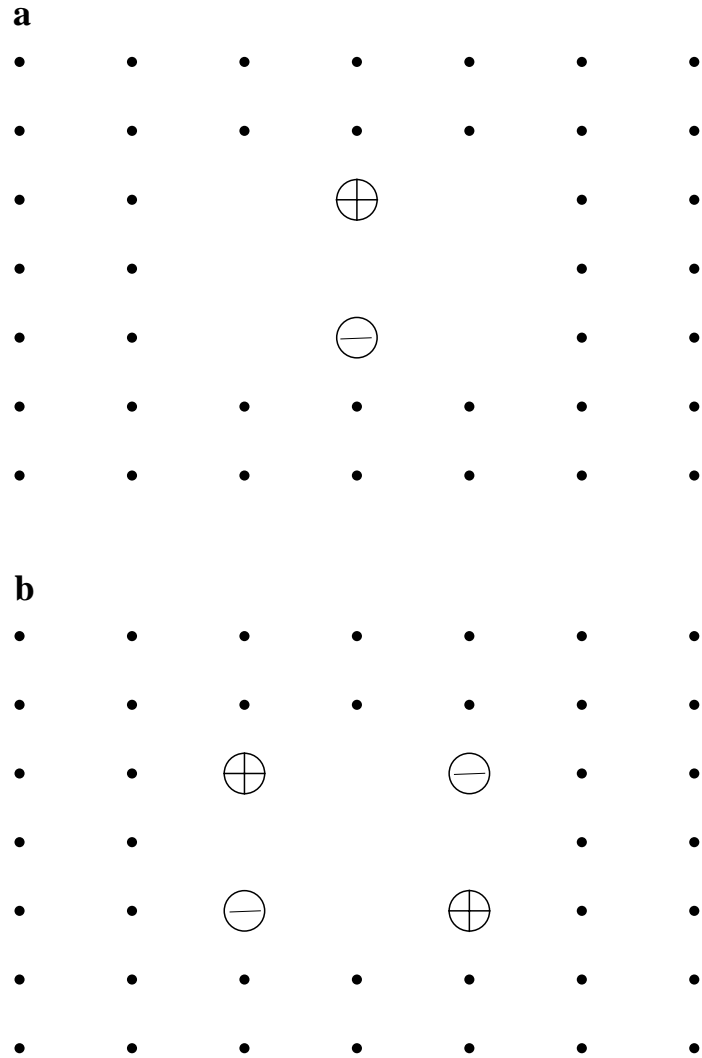}
\end{center}
\caption{Dipole (a) and quadrupole (b) sources. Points indicate the equivalent source positions. Circles represent the monopole sources placed at the points of the small cube. Symbols + and –- indicate the signs of the monopole source strengths. }%
\label{fig6}%
\end{figure*}

Thus, we obtained a strongly underdetermined system: there are only 5
equations for finding 316 unknown $A_{n}$. Besides, intuition suggests that in
order to reconstruct the radiator field, receivers should surround it on all sides.

Fortunately, the underdeterminedness of the problem can be significantly reduced without additional measurements. Consider the rotations of our dipole by 90$^{\circ}$, 180$^{\circ}$, and 270$^{\circ}$ about each of the $x$, $y$ and $z$ axes. Since at each rotation the points $\mathbf{r}_{+}$ and $\mathbf{r}_{-}$ fall into other points of the small cube, for each of the ten
resulting dipole positions (original position and 9 rotations) we can easily find the corresponding vector of field amplitudes on the receivers $\mathbf{\tilde{u}}=\mathbf{\tilde{u}}_{j}$, $j=1,\ldots,10$. 

Note that the rotation of the dipole allows us to get by with a smaller number of receivers.
Instead of using receivers surrounding the dipole from all sides,
we alternately turn its different sides to the 5 existing receivers.

In accordance with our assumptions, formulated in Sec. \ref{sub:esm-tank}, the
dipole at any turn is associated with the same equivalent sources that should
be rotated with it. For each of the turns described above, the equivalent
sources only change places: the equivalent source from the point
$\mathbf{\tilde{r}}_{n}$ falls into the point $\mathbf{\tilde{r}}_{m}$,
representing the position of another equivalent source. Then, the source
strength at the point $\mathbf{\tilde{r}}_{m}$ becomes equal to $A_{n}$. This
means that the rotation of the dipole results in a permutation of the elements
of the vector $\mathbf{A}$. For the j-th dipole position, the system of
equations (\ref{uGA-matr}) takes the form%
\begin{equation}
\mathbf{\tilde{u}}_{j}=\mathbf{\tilde{G}T}_{j}\mathbf{A,} \label{utl}%
\end{equation}
where $\mathbf{T}_{j}$ is a permutation matrix of size $316\times316$,
describing the permutation of the elements of vector $\mathbf{A}$ caused by
the rotation.

Combining 10 vectors $\mathbf{\tilde{u}}_{j}$ into one vector and 10 matrices
$\mathbf{\tilde{G}T}_{j}$ into one matrix,
\[
\mathbf{\tilde{u}}_{c}=\left(
\begin{array}
[c]{c}%
\mathbf{\tilde{u}}_{1}\\
\vdots\\
\mathbf{\tilde{u}}_{10}%
\end{array}
\right)  ,\;\mathbf{\tilde{G}}_{c}=\left(
\begin{array}
[c]{c}%
\mathbf{\tilde{G}T}_{1}\\
\vdots\\
\mathbf{\tilde{G}T}_{10}%
\end{array}
\right)  ,
\]
we get a system of 50 equations for 316 unknowns $A_{n}$:%
\begin{equation}
\mathbf{\tilde{u}}_{c}=\mathbf{\tilde{G}}_{c}\mathbf{A.} \label{uc-Gc}%
\end{equation}

So, the rotations of the dipole and equivalent sources at the angles indicated
above allowed us to increase the number of equations by a factor
of 10 without increasing the number of receivers and without additional measurements.

It should be noted that there are 24 different rotations of the cube, in which
its points exchange places: the lower side of the cube can be selected in 6
ways and this\ side can then be rotated into 4 different positions. We use
only 10 rotations since this is sufficient to reconstruct the fields of the
simplest sources considered in this paper.

To solve the system (\ref{uc-Gc}), we use the method described in Sec. \ref{sub:free}. The desired estimate of the vector $\mathbf{A}$ is given by
the formula (\ref{At}), in which $\gamma_{l}$ are the singular numbers of the
matrix $\mathbf{\tilde{G}}_{c}$ (see Fig. 7), $\pmb{\xi}_{l}$ and
$\pmb{\eta}_{l}$ are the singular vectors. For each value of $L_{1}$ from 1
to 50, Eq. (\ref{At}) gives an approximate solution of the system
(\ref{uc-Gc}). A quantitative estimate of the accuracy of each solution is
given by the parameter $\mu$ determined by Eq. (\ref{mu}) with $\mathbf{u}%
=\mathbf{\tilde{u}}_{с }$ and $\mathbf{\hat{u}}=\mathbf{\tilde{G}}%
_{c}\mathbf{\hat{A}}$. The dependence of $\mu$ on $L_{1}$ is shown in the
upper panel of Fig. 8.

\begin{figure*}
\begin{center}
\includegraphics[
width=1\linewidth]{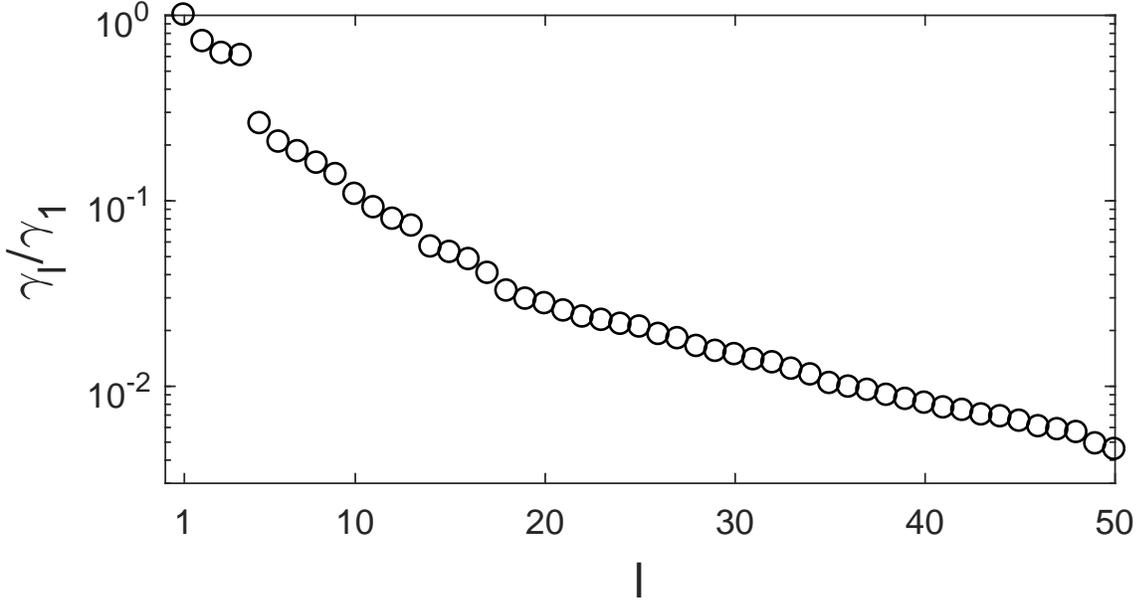}
\end{center}
\caption{Singular values of matrix $\mathbf{\tilde{G}}_c$ divided by the greatest of them.}%
\label{fig7}%
\end{figure*}

\begin{figure*}
\begin{center}
\includegraphics[
width=1\linewidth]{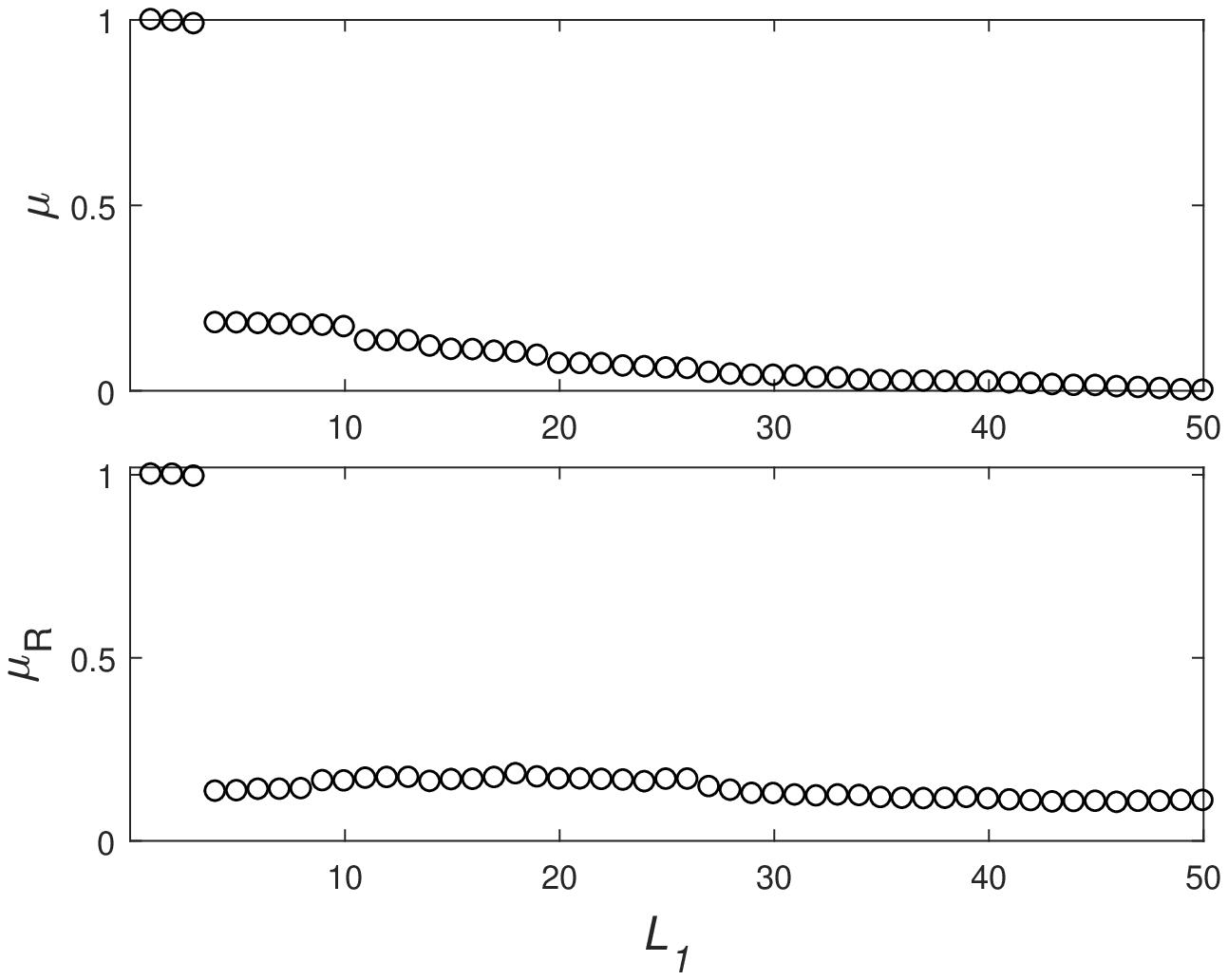}
\end{center}
\caption{Upper panel: parameter $\mu$ quantitatively characterizing the closeness of the measured and reconstructed field at the collocation points as a function of $L_1$, the number of singular values taken into account. Lower panel: similar dependence for parameter $\mu_R$ quantitatively characterizing the accuracy of the reconstructed dipole field in free space.}%
\label{fig8}%
\end{figure*}

Substituting the found source strengths $A_{n}$ into Eq. (\ref{uGA}) gives an
approximate expression for the field excited by the dipole source in free
space $u\left(  \mathbf{r}\right)  $. To estimate the reconstruction accuracy,
we compare the values of the function found $u\left(  \mathbf{r}\right)  $ on
a spherical surface of radius $R=1$ m in free space with the values on the
same surface of the function $u_{e}\left(  \mathbf{r}\right)  =G\left(
\mathbf{r},\mathbf{r}_{+}\right)  -$ $G\left(  \mathbf{r},\mathbf{r}%
_{+}\right)  $ representing the exact expression of the field excited by our
dipole. The points of this surface are given by the relations $x=R\cos
\varphi\sin\theta$, $y=R\sin\phi\sin\theta$, $z=R\cos\theta$, where $\phi$ and
$\theta$ are azimuthal and polar angles, respectively. Choosing discrete
angles $\phi_{m}$, $m=1,\ldots,100$, and $\theta_{n}$, $n=1,\ldots,90$
uniformly sampling the intervals $0\leq\phi\leq2\pi$, and $0\leq\theta\leq\pi
$, respectively, we get sampling points $\mathbf{r}_{mn}$~on the selected
surface. By analogy with (\ref{mu}) as a quantitative measure of the proximity
of the $u\left(  \mathbf{r}\right)  $ and $u_{e}\left(  \mathbf{r}\right)  $
we take
\begin{equation}
\mu_{R}=\left(  \frac{\sum_{m,n}\left\vert u\left(  r_{mn}\right)
-u_{e}\left(  r_{mn}\right)  \right\vert ^{2}}{\sum_{m,n}\left\vert
u_{e}\left(  r_{mn}\right)  \right\vert ^{2}}\right)  ^{1/2}, \label{mu-R}%
\end{equation}
where the summation goes over all sampling points. The lower panel in Fig. 8 shows
the values of $\mu_{R}$ for all possible values of $L_{1}$. As we see, for all
$L_{1}\geq4$ both $\mu$ and $\mu_{R}$ are small compared to the unity. This
means that with such $L_{1}$, the inverse problem is solved with satisfactory accuracy.

Figure 9 shows the amplitudes $\left\vert u_{e}\left(  \mathbf{r}\right)
\right\vert $ (upper panel) and $\left\vert u\left(  \mathbf{r}\right)
\right\vert $ (lower panel) on the spherical surface as functions of the
azimuthal and polar angles. These plots actually represent the exact and
reconstructed directivity patterns of the dipole source. Amplitude $\left\vert
u\left(  \mathbf{r}\right)  \right\vert $, whose distribution on the spherical
surface is shown in the lower panel, is calculated with $L_{1}=8$. The
distributions of amplitudes $\left\vert u\left(  \mathbf{r}\right)
\right\vert $ for all $L_{1}$ from interval $4\leq L_{1}\leq50$ look likewise.

\begin{figure*}
\begin{center}
\includegraphics[
width=1\linewidth]{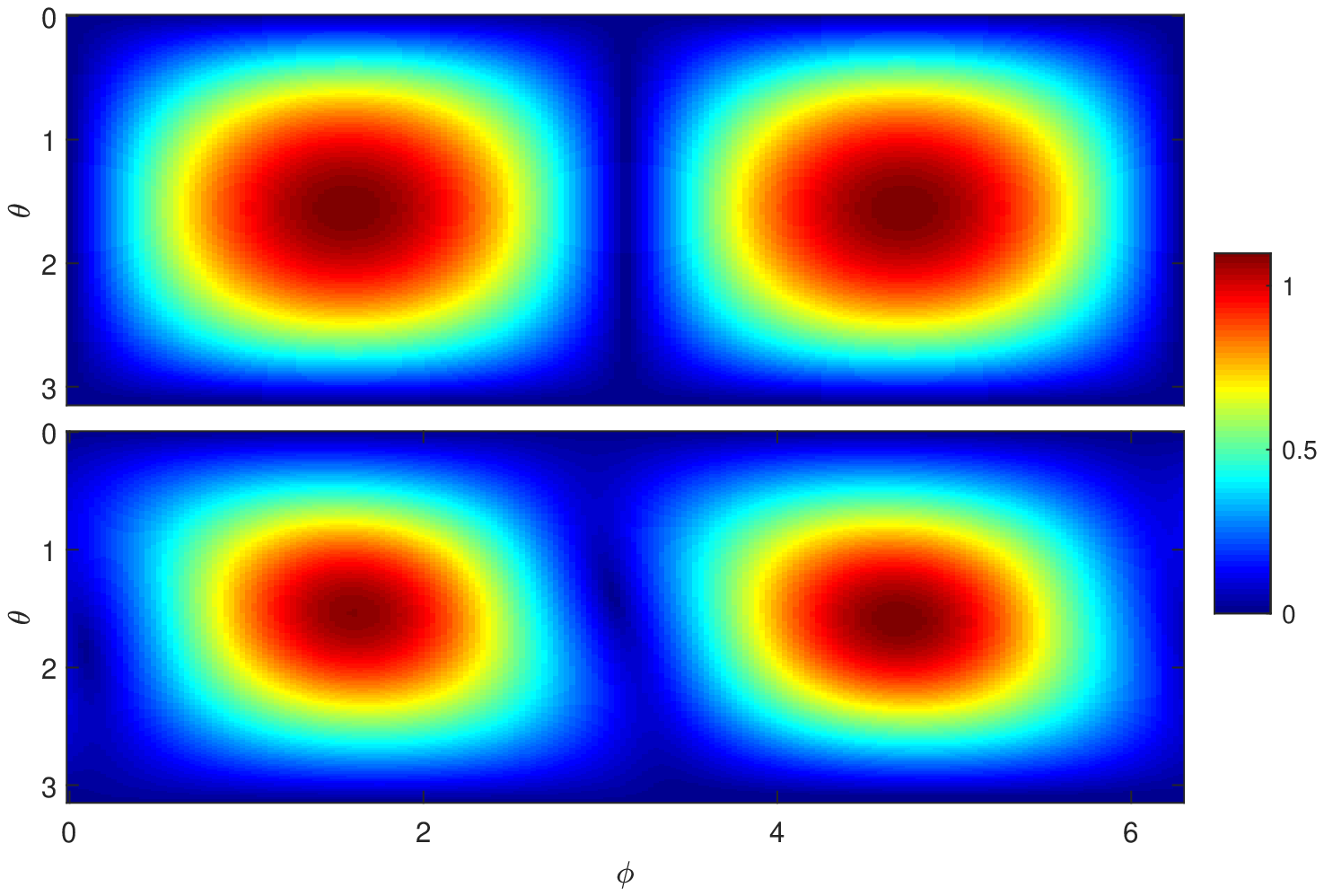}
\end{center}
\caption{Field amplitude at 1 m from the dipole center in free space as a function of azimuth angle $\phi$ and polar angle $\theta$. Upper panel: exact analytic solution. Lower panel: result of the reconstruction with $L_1$ = 8.}%
\label{fig9}%
\end{figure*}

It is worth noting, that despite the proximity of the functions $u\left(
\mathbf{r}\right)  $ found for all $L_{1}$ from the above interval, the source
strengths corresponding to different $L_{1}$ are generally strongly different.
To illustrate this statement, Fig. 10 shows the absolute values of source
strength  found with $L_{1}$ = 8 (left panel) and $L_{1}$ = 16 (right panel).
The size of a point whose center is located at the position of the $n$-th
equivalent source is proportional to $\left\vert A_{n}\right\vert $. 

\begin{figure*}
\begin{center}
\includegraphics[
width=1\linewidth]{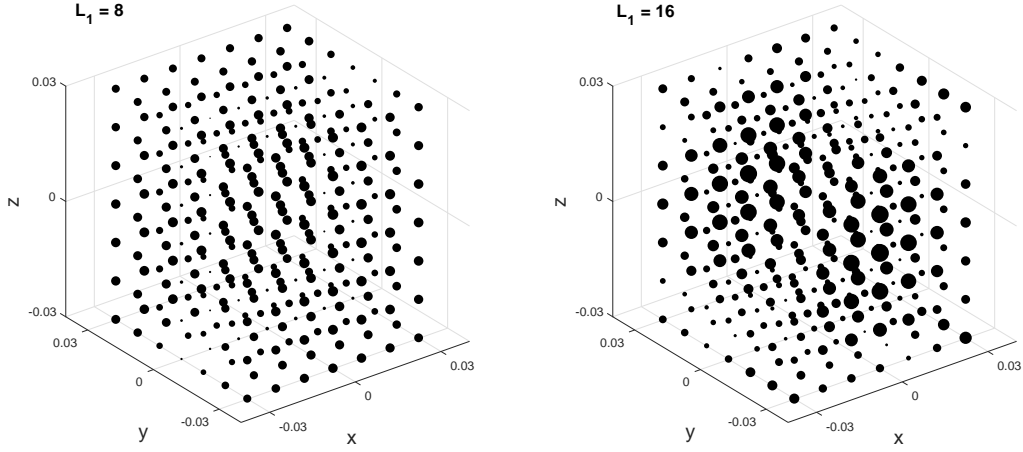}
\end{center}
\caption{Equivalent source strengths found by solving the system (22) with the account of 8 (left panel) and 16 (right panel) singular values and vectors.}%
\label{fig10}%
\end{figure*}

Similarly, it is possible to reconstruct the field of a quadrupole in free
space. Its field in the tank can be synthesized from the fields emitted by the
etalon monopole from four points of the small cube shown in Fig. 6b. In this
example, the four points are located in the $x-y$ plane. The
inverse problem again reduces to solving the system (\ref{uc-Gc}) with the same
matrix $\mathbf{\tilde{G}}_{c}$ but different $\mathbf{\tilde{u}}_{c}$. As in the case of dipole, the source
strengths $A_{n}$, found by the formula (\ref{At}), are to be substituted in
Eq. (\ref{uGA}). To estimate the accuracy of the reconstruction, the same
parameters $\mu$ and $\mu_{R}$ are used as for the dipole. Figure 11 presents
the dependencies of these parameters on $L_{1}$, the number of singular values
and vectors taken into account. As we see, for the quadrupole these parameters
are small for $L_{1}\geq8$. Figure 12 shows the amplitude distributions of the
quadrupole field on the spherical surface of radius $R$ = 1 cm calculated
using the exact analytical formula (upper panel) and reconstructed by solving
the inverse problem with $L_{1}$ = 8 (lower panel). As in the case of the
dipole, these directivity patterns coincide well. The same is true for all
$L_{1}$ from interval $8\leq L_{1}\leq50$ (not shown). 

\begin{figure*}
\begin{center}
\includegraphics[
width=1\linewidth]{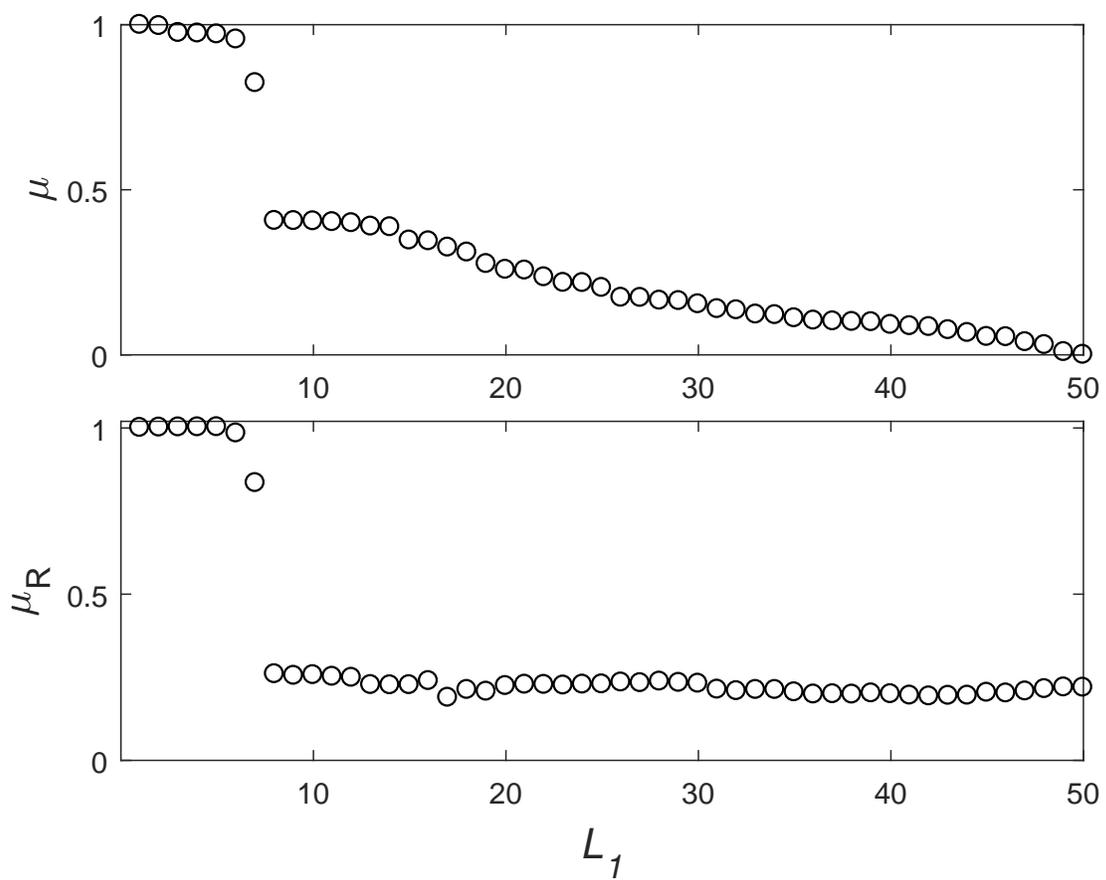}
\end{center}
\caption{The same as in Fig. 8, but for the quadrupole in the plane $x-y$.}%
\label{fig11}%
\end{figure*}

\begin{figure*}
\begin{center}
\includegraphics[
width=1\linewidth]{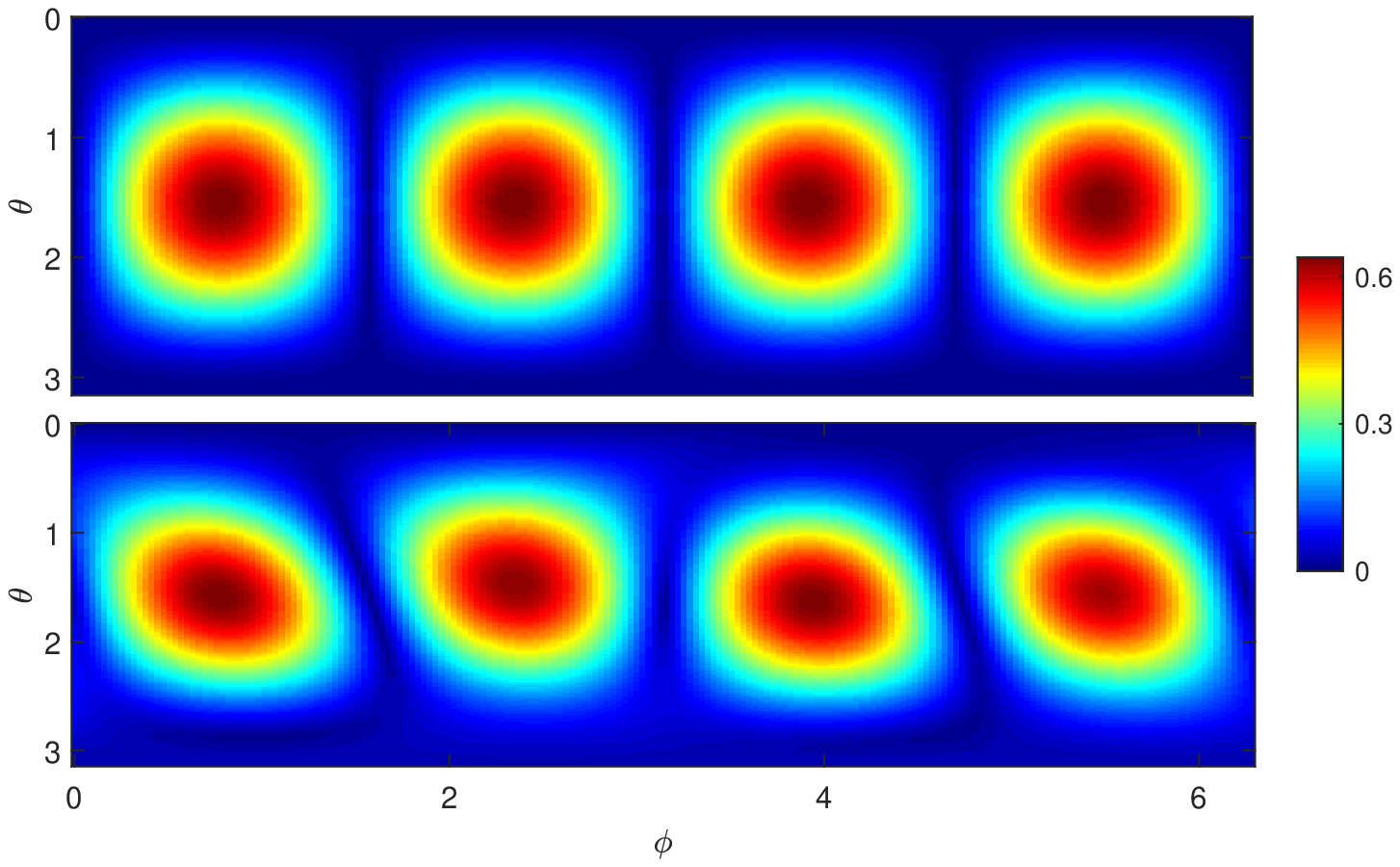}
\end{center}
\caption{The same as in Fig. 9, but for the quadrupole in the plane $x-y$.
Upper panel: exact analytic solution. Lower panel: result of the reconstruction with $L_1$ = 8.}%
\label{fig12}%
\end{figure*}

Similar results were obtained when reconstructing the field of a quadrupole
located in the $x-z$ plane. Exact and reconstructed (with $L_{1}$ =
8) directivity patterns of this quadrupole are presented in Fig. 13.

\begin{figure*}
\begin{center}
\includegraphics[
width=1\linewidth]{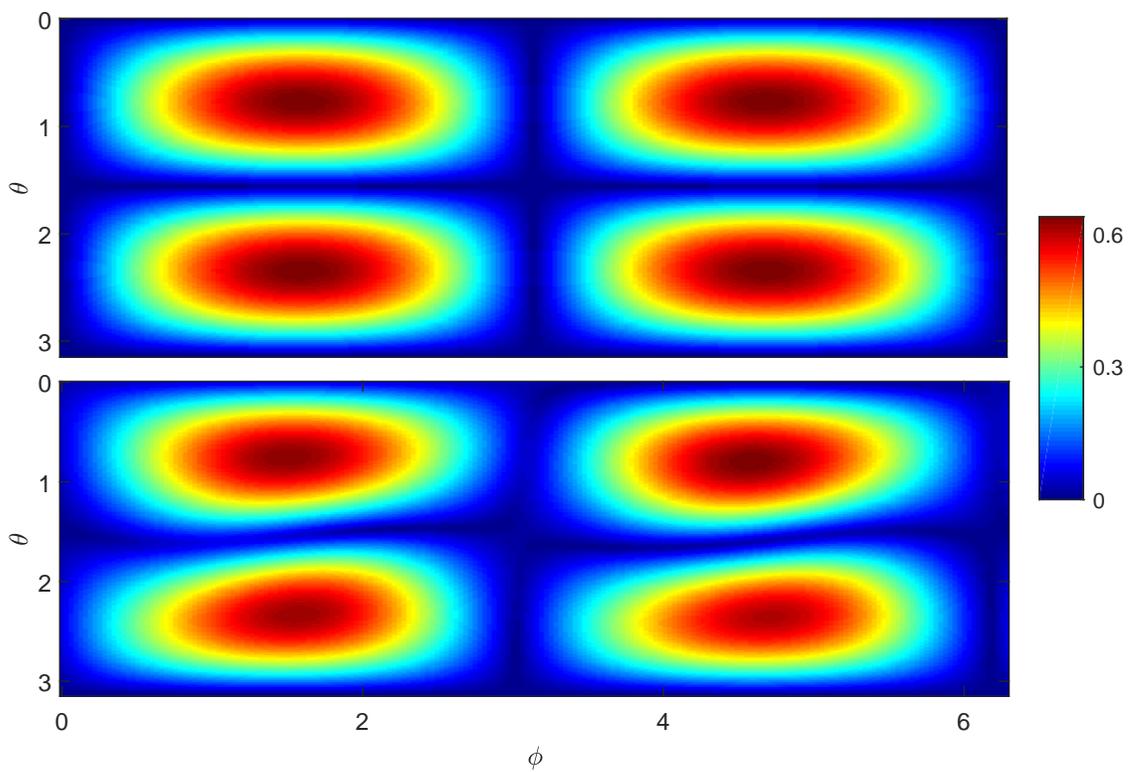}
\end{center}
\caption{The same as in Fig. 9, but for the quadrupole in the plane $x-z$.
Upper panel: exact analytic solution. Lower panel: result of the reconstruction with $L_1$ = 8.}%
\label{fig13}%
\end{figure*}

We also considered the field excited by a monopole located at one of the
points of the small cube (not shown). Its field in free space is successfully
reconstructed with $L_{1}\geq2$.

\section{Conclusion \label{sec:conclusion}}

The paper shows that the use of the equivalent source method makes it possible
to reconstruct the field of a radiator in free space from the measurements of
its field in a tank with reflecting boundaries. The approach under
consideration allows the free-space calibration of the radiator in the tank
avoiding the isolation of signals arriving at the receivers without boundary
reflections. A key role in solving the inverse problem plays the tank
calibration. This procedure gives the necessary values of the tank Green
function, which accounts for multiple reflections of sound waves from the boundaries.

To test the feasibility of the proposed approach, a tank experiment was
carried out. Analysis of the experimental data confirmed our assumption that
equivalent sources reproducing the field of the radiator in free space
reproduce the field of the same radiator in the tank. It was also demonstrated
that the proposed method allows one reconstructing the fields of an acoustic
monopole, dipole, and quadrupole in free space by measuring the fields of
these sources in the tank.

As noted earlier, the weak point of the method of equivalent sources and,
accordingly, the weak point of our approach is the lack of general rules for
choosing the necessary number of equivalent sources and their positions. In
our case, this difficulty is reinforced by the requirement that the same
sources provide a sufficiently accurate approximation of the radiator field by
the sum (\ref{uGA}) in free space and by the sum (\ref{uGAt}) in the tank. In
section \ref{sub:esm-tank} conditions (i) - (iv) are listed, which we consider
necessary for this. However, these conditions are formulated in too general
terms and they need to be made more precise and concrete.

Finally, it should be noted that the equivalent source method is used to
analyze the scattered fields \cite{Johnson1998,Gounot2011}. The approach
considered in this paper can also be applied for this purpose. With it, one
can reconstruct the field scattered on the body in free space from the
measurements of the field scattered on the same body in the tank. To solve this problem, the scatterer together with the radiator can be considered as one single sound source. Then its field in free space can be reconstructed using the discussed approach.

\section*{Acknowledgements}

The research was carried out within the state assignment of IAP RAS 
(Project \#00035-2014-0011). We are grateful to Dr. L.Ya. Lyubavin for valuable discussions.

\end{document}